\input  phyzzx
\input epsf
\overfullrule=0pt
\hsize=6.5truein
\vsize=9.0truein
\voffset=-0.1truein
\hoffset=-0.1truein

%
%

\def\IC{{\ \hbox{{\rm I}\kern-.6em\hbox{\bf C}}}}
\def\IR{{\hbox{{\rm I}\kern-.2em\hbox{\rm R}}}}
\def\IZ{{\hbox{{\rm Z}\kern-.4em\hbox{\rm Z}}}}

\def\sIR{{\hbox{{\sevenrm I}\kern-.2em\hbox{\sevenrm R}}}}

%
%
\hyphenation{Min-kow-ski}

\rightline{SU-ITP-97-12}
\rightline{April  1997}
\rightline{hep-th/9704086}

\vfill

%
%
\title{A Proposal for the Wrapped Transverse Five Brane in M(atrix) Theory}

\vfill

%
%

\author{Edi Halyo\foot{halyo@dormouse.stanford.edu}}

\bigskip

\address{Department of Physics \break Stanford University \break
 Stanford, CA 94305-4060}

\vfill

%
%

We propose a state in $5+1$ super Yang--Mills theory which corresponds to the
wrapped transverse five brane of  M(atrix) theory on $T^5$. This state is a
magnetic flux quantized in units of $1/g_6^2$
through a plane defined by one side of the box and a new
direction which is not manifest.

\vfill\endpage

%
%

\REF\bfss{T. Banks, W. Fischler, S. Shenker and L. Susskind, hep-th/9610043.}
\REF\sca{O. Aharony and M. Berkooz, hep-th/9611215, G. Lifschytz and S. Mathur,
hep-th/9612087; G. Lifschytz, hep-th/ 9612123.}
\REF\str{L. Motl, hep-th/9701025; T. Banks and N. Seiberg, hep-th/9702187; R.
Dijkgraaf, E. Verlinde and H. Verlinde, hep-th/9703030; Y. Imamura,
hep-th/9703077; T. Banks and L. Motl, hep-th/9703218; D. Lowe, hep-th/9704041.}
\REF\len{S. Sethi and L. Susskind, hep-th/9702101; L. Susskind, hep-th/9611164;
O.J. Ganor, S. Ramgoolam and W. Taylor, hep-th/9611202.}
\REF\bd{M. Berkooz and M. R. Douglas, hep-th/9610236.}
\REF\bss{T. Banks, N. Seiberg and S. Shenker, hep-th/9612157.}
\REF\fbr{A. Fayazuddin and D. Smith, hep-th/9703208; G. Lifschytz,
hep-th/9703201.}
\REF\wati{W. Taylor,hep-th/9611042.}
\REF\fhrs{W. Fischler, E. Halyo, A. Rajaraman and L. Susskind, hep-th/9703102.}
\REF\roz{M. Rozali, hep-th/9702136.}


%
%

Recently, there has been great interest in the matrix model formulation of M
theory[\bfss].
In this framework, M theory in the light--cone coordinates is described by an
infinite number of zero
branes and their interactions due to open strings stretched between them. This
is
a $0+1$ dimensional $U(N)$ super Yang--Mills (SYM) field theory  with sixteen
supersymmetries which is obtained by the dimensional reduction of  $D=10$ SYM
theory with ${\cal N}=1$ supersymmetry.
By now, the matrix model has passed a number of consistency checks such as
long range interactions between gravitons, membranes etc.[\sca], the
formulation of
different types of strings[\str] and the presence of U dualities[\len].
Fundamental states of M theory such as
gravitons and membranes have been easily identified but the five branes of the
theory turned
out to be more elusive. The longitudinal five brane has been constucted as a
background [\bd] but the transverse five brane seems to be absent in the matrix
model[\bss]. (However, see [\fbr]
for work related to the M(atrix) theory five brane.)

When M theory is compactified on $T^5$, the transverse five brane can be
wrapped
completely on the five torus. In this case,
one expects the transverse five brane to appear in the spectrum of the matrix
model
on the same footing as the other pointlike BPS states. Since
matrix model on $T^d$ is described by $U(N)$ SYM theory in a $d$ dimensional
box[\bfss,\wati], the transverse
five brane must be one of the BPS states of this theory. This theory has
sixteen BPS
states and fifteen of them have been identified in terms of the Abelian
electric and magnetic
fluxes of the SYM theory[\fhrs].
In this note, we propose a candidate for the BPS state of the SYM theory which
corresponds to the  transverse five brane wrapped on $T^5$. This state is
simply
a magnetic field which is quantized in units of $1/g^2_6$, where $g_6$ is the
$5+1$
dimensional SYM
coupling with dimension of length. The direction of the magnetic field (or
flux) is given by one side of the box and a new dimension which is not
manifest. This new
dimension is the one which opens up in a $4+1$ dimensional SYM theory as one
increases the gauge coupling $g_5^2$[\roz]. It is in some sense conjugate to
the instanton (or soliton) number of the $4+1$ SYM theory.
The energy of this state SYM
theory matches the light--cone energy
which corresponds to the wrapped transverse five brane.

In the following, we first describe the $5+1$ dimensional $U(N)$ SYM theory and
its manifest fifiteen BPS states which are given in terms of the electric and
magnetic fluxes.
We then describe the BPS state which corresponds to the transverse five brane
and calculate its mass. We end with a short discussion of our results.

As mentioned above, matrix model on $T^5$ (with compact dimensions of length
$L_1,L_2,L_3$ $,L_4,L_5$)
is described by an $U(N)$ SYM theory in a five dimensional box with sides
(parametrized by $\sigma_i, \quad i=1,\ldots,5$)
[\fhrs]
$$\Sigma_i={(2\pi)^3 \ell_{11}^3 \over {RL_i}} \eqno(1)$$
and coupling constant
$$g_6^2={(2\pi)^9 \ell_{11}^9 \over{R^2 L_1 L_2 L_3 L_4 L_5}} \eqno(2)$$
with dimension of area. The Lagrangian of the model is given by
$${\cal L}= \int^V d^5 \sigma \quad Tr(-{1\over 4g_6^2}F_{\mu
\nu}^2+(D_{\mu}\phi_i)^2+g_6^2[\phi_i,\phi_j]^2+fermionic \quad terms)
\eqno(3)$$
where $V=\Sigma_1 \Sigma_2 \Sigma_3 \Sigma_4 \Sigma_5$ is the volume of the
box,
$\mu=0,\ldots,5$, $i=6,\ldots,9$ and all fields are
in the adjoint representation of $U(N)$.
M(atrix) theory on $T^5$ has sixteen pointlike BPS states. Five of these are
Kaluza--Klein or momentum states with mass
$$M={2\pi \over L_i} \eqno(4)$$	
and light--cone energy $H=M^2/2p_{11}=M^2 R/2N$
$$H={(2\pi)^2 R \over{2N L_i^2}} \eqno(5)$$
In the SYM theory, these are described by $U(1)$ electric fluxes which satisfy
$$\epsilon_{ijklm}E_i \Sigma_j \Sigma_k \Sigma_l \Sigma_m={2\pi n_i \over
N}\eqno(6)$$
Using
$$H_e={N \over 2}g_6^2 E^2 \Sigma_1 \Sigma_2 \Sigma_3 \Sigma_4 \Sigma_5
\eqno(7)$$
we find precisely the light--cone energy in eq. (5).
In addition, there are ten BPS states which correspond to membranes wrapped
on the ten two tori of $T^5$ with mass
(the membrane tension is 1$/(2\pi)^2 \ell_{11}^3$)
$$M={L_i L_j \over{(2\pi)^2 \ell_{11}^3}} \eqno(8)$$
and light--cone energy
$$H={(L_i L_j)^2 R\over {(2\pi)^4 \ell_{11}^6 2N}} \eqno(9)$$
These are described by the ten $U(1)$ magnetic fluxes which satisfy the
quantization
condition
$$B_{ij} \Sigma_i \Sigma_j={2\pi n_{ij} \over N} \eqno(10)$$
Using 
$$H_m={N \over 2}g_6^{-2} B^2 \Sigma_1 \Sigma_2 \Sigma_3 \Sigma_4 \Sigma_5
\eqno(11)$$
we find precisely the light--cone energy in eq. (9).

The sixteenth BPS state corresponds to the wrapped transverse five brane
with mass
(the five brane tension is $1/(2\pi)^5 \ell_{11}^6$)
$$M= {L_1 L_2 L_3 L_4 L_5 \over (2\pi)^5 \ell_{11}^6} \eqno(12)$$
This state could not be identified in the $5+1$ SYM theory in terms of the
usual Abelian fluxes[\fhrs].
We propose that the transverse five brane is described by a magnetic
flux quantized in units of $1/g_6^2$ (similar to the other magnetic fluxes
quantized in units of inverse the area of the faces of the five
dimensional box), i.e.
$$Bg_6^2={2\pi n \over N} \eqno(13)$$
We also need to specify the direction of this magnetic flux. This can be done
by realizing that
$$g_6^2=g_5^2 \Sigma_5=\Sigma \Sigma_5 \eqno(14)$$
where $g_5^2=(2\pi)^6\ell_{11}^6/R L_1 L_2 L_3 L_4$ is the coupling constant of
the $U(N)$ SYM in a four dimensional box. It has been argued that $g_5^2$
corresponds to a new dimension of this box (parametrized by $\sigma$ and with
size $\Sigma$) which opens up as the coupling increases[\roz]. The reason for
this is as follows. The $4+1$ dimensional SYM theory has instantons (or
solitons) which correspond to the longitudinal five branes of M theory with
energy $n/g_5^2$. Since $g_5^2=\Sigma$ has dimension of length, these can be
interpreted as momentum modes of a new dimension with size $\Sigma$.
Thus the magnetic field in eq. (13) is in the direction $\sigma \sigma_5$ which
is not manifest in the five dimansional box.
Substituting this into eq. (11) we get
$$\eqalignno{H_m&={N \over 2}{(2\pi)^2\over g_6^2} {1\over g_6^4} \Sigma_1
\Sigma_2
\Sigma_3 \Sigma_
4 \Sigma_5 &(15a) \cr
&={R (L_1 L_2 L_3 L_4 L_5)^2 \over {(2\pi)^{10}\ell_{11}^{12}} }
&(15b)} $$
This reproduces the correct mass for the wrapped five brane.
The above description of the transverse five brane is in some sense similar to
the description of wrapped transverse membranes. These are described by
magnetic fluxes in eq. (10) whereas longitudinal membranes wrapped on $RL_i$
are given by momentum modes (or photons)
in the box, i.e. states with $p_i=n/\Sigma_i$. On the other hand, the
longitudinal five brane is described by the momentum mode $p=n/\Sigma$. Thus,
in analogy,  we expect that the transverse five brane is given by the magnetic
flux
$$B_{\sigma \sigma_5} \Sigma \Sigma_5=2\pi n \eqno(16)$$
which is precisely eq. (13).
A possible gauge field configuration which describes the transverse five brane
is
$$A_5={2\pi \over {g_6^2 N}} \sigma I_{N \times N} \eqno(17)$$
Note that among all the BPS states only the longitudinal and transverse
five branes are described in terms of the new direction $\sigma$.

The sixteen BPS states are in the spinor representation of the U duality group
$SO(5,5;Z)$.
However, in the five dimensional box only the $SL(5;Z)$ part of this is
manifest as a
geometrical symmetry. Under $SL(5;Z)$ the sixteen BPS states decompose as a
$10$ (the wrapped membranes), a $5$ (the Kaluza--Klein states) and a singlet
which is the wrapped
five brane.  The five brane is a singlet because the factor
$\Sigma \Sigma_5=\ell_{11}^9/R^2 L_1 L_2 L_3 L_4 L_5$ which defines it is
invariant under permutations of $\Sigma_i$ (or $L_i$). The electric and
magnetic fluxes transform among themselves and form  the 5 and 10
representations. The transformations
$\Sigma \leftrightarrow \Sigma_i$ exchange the five brane with one of the
membranes.
These are some of the generators of $SO(5,5;Z)$ which are not in $SL(5;Z)$.

In this note we proposed a state in the $5+1$ SYM theory as a candidate for the
wrapped transverse five brane of M theory. This is simply a magnetic flux
through a plane defined by one side of the box and a new dimension which is not
manifest. The interpretation of this dimension which is roughly conjugate to
instanton number of $4+1$ SYM theory is not clear. For this reason, it is not
possible to write down the
configuration of the five brane only in terms of the variables of the SYM
theory in a
five dimensional box. This may also be the reason for the absence of the five
brane
central charge in the matrix model. Clearly more needs to be done to show that
this state in fact describes the transverse five brane. First and foremost a
better understanding of the $\sigma$ direction in terms of the five dimensional
box
variables is needed. This will enable us to write down a configuration which
corresponds to the five brane. Once the configuration is known, one can try to
show that membranes can end on these configurations and/or calculate the
Berry's phase
obtained by taking a membrane around the five brane.

\bigskip
\centerline{\bf Acknowledgements}

We would like to thank  Moshe Rozali and Lenny Susskind for discussions.
\vfill
\endpage

\refout
\end
\bye